\documentclass[10pt,twocolumn,showpacs,prb]{revtex4}
\usepackage{graphicx}

\begin{document}

\title{Simulation of  adsorbate-induced faceting on curved surfaces.}
\author{ Daniel Niewieczerza{\l} and Czes\l{}aw Oleksy\footnote{Corresponding author.\\
Electronic address: oleksy@ift.uni.wroc.pl}}

\affiliation{ Institute of Theoretical Physics, University of
Wroc{\l}aw,
 Plac Maksa Borna 9, 50-204 Wroc{\l}aw, Poland }
%\date{\today}

\begin{abstract}
A  simple solid-on-solid model, proposed earlier to describe
overlayer-induced faceting of bcc(111) surface, is applied to
faceting of  curved surfaces covered by an adsorbate monolayer.
Surfaces studied in this paper are formed by a part of sphere
around the [111] pole. Results of Monte Carlo simulations  show
that the morphology of a faceted surface depends on the annealing
temperature. At an initial stage the surface around the $[111]$
pole consists of 3-sided pyramids and step-like facets, then
step-like facets dominate and their number decreases with
temperature, finally a single big pyramid is formed. It is shown
that there is a reversible phase transition at which a faceted
surface transforms to an almost spherical one. It is found that
the temperature of this phase transition is an increasing function
of the surface curvature. Simulation results show that
measurements of high temperature properties performed directly and
after fast cooling down to a low temperature lead to different
results.

\end{abstract}


\maketitle
\section{Introduction}

It has been recently demonstrated that ultrathin  metal films
induce faceting of bcc(111) surfaces. Atomically rough W(111) and
Mo(111) surfaces, when covered by a single physical monolayer of
certain metals (Pd, Rh, Ir, Pt, Au) and annealed  to  $T>700$K,
undergo massive reconstruction from a planar morphology  to a
microscopically faceted surface
\cite{mad94,mad95,song95,mad96,mad97,mad99a,mad99b}. In most of
investigated bcc(111) surfaces, the faceted morphology comprises
3-sided pyramids with $\{211\}$ facets and facet sizes ranging
from $\sim 3$ to 100 nm. It has been shown that the metal that
induces the faceting acts like a surfactant and remains on the
surface during the faceting transformation \cite{mad99b}.

Suggestions that faceting in these systems is thermodynamically
favorable have been confirmed by first-principles
calculations\cite{mad99b,leung97,leung98} performed for (111),
(211), and (110)  surfaces of Mo and W. Results  show lowering of
the surface energy as surfaces are being covered by metal
overlayer (e.g. Pd, Pt, Au). Moreover, at coverage of one physical
monolayer, the bcc(111) surface becomes unstable against faceting,
i.e. the system gains energy by transition from (111) to $\{211\}$
orientation.

Interesting results have been obtained for faceting on curved
surfaces of W and Mo in FIM (field ion microscopy)
\cite{szczep02,szczep04,szczep05} and FEM (field emission
microscopy) experiments\cite{pelhos99,antczak01,antczak03}.
Initially a spherical surface comprises macroscopic crystal
facets. After depositing a metal film (e.g. Pd, Pt) or oxygen
layer and annealing at elevated temperatures, an increase of the
macroscopic \{110\} and \{211\} crystal facets is observed. An
unexpected change in the faceted morphology occurs near the
$\langle111\rangle$ poles, where step-like \{211\} microfacets are
formed \cite{szczep02,szczep04}. The number of step-like facets
decreases with the annealing temperature, and in some cases, a
single \{211\} pyramid around the $[111]$ pole is
observed\cite{szczep04}. Moreover, occurrence  of new facets
\{123\}, \{178\} in Pt on W and Pd on Mo  has been reported in
papers on FEM \cite{antczak01,antczak03}.

In theoretical studies of complicated surface problems (e.g.
roughening transition, surface reconstruction, surface growth,
surface phase transitions)   simple models  like lattice gas
models or  solid-on-solid (SOS) models are applied
\cite{levi97,sasaki94,beijeren77,beijeren95,tosat95,tosat96,bartelt94,kaseno97,salanon88,nijs89,jaszczak97}.
In our earlier paper \cite{oleksy} we have proposed a  SOS model
to study the adsorbate-induced faceting of the bcc(111) crystal
surface at constant coverage. Monte Carlo simulation results show
formation of pyramidal facets in accordance with experimental
observations. Moreover, the model describes a reversible phase
transition from a faceted surface to a disordered (111) surface.
Such a phase transition has been found in the Pd/Mo(111) system
\cite{song95}. In this paper we study faceting of a curved bcc
crystal surface by using the SOS model. We focus on changes in
surface morphology in the vicinity of the $[111]$ pole, where
step-like \{211\} facets were observed in FIM experiments. The
paper is organized  as follows. In  Sec.~\ref{s2} the SOS model
for a curved surface of a bcc crystal  is described. Results of
Monte Carlo simulations, including changes in the surface
morphology with the annealing temperature, and a reversible phase
transition,  are presented in Sec.~\ref{s_MC}.

\section{The SOS model}\label{s2}

\begin{figure}
\centering
\includegraphics[width=7.5cm]{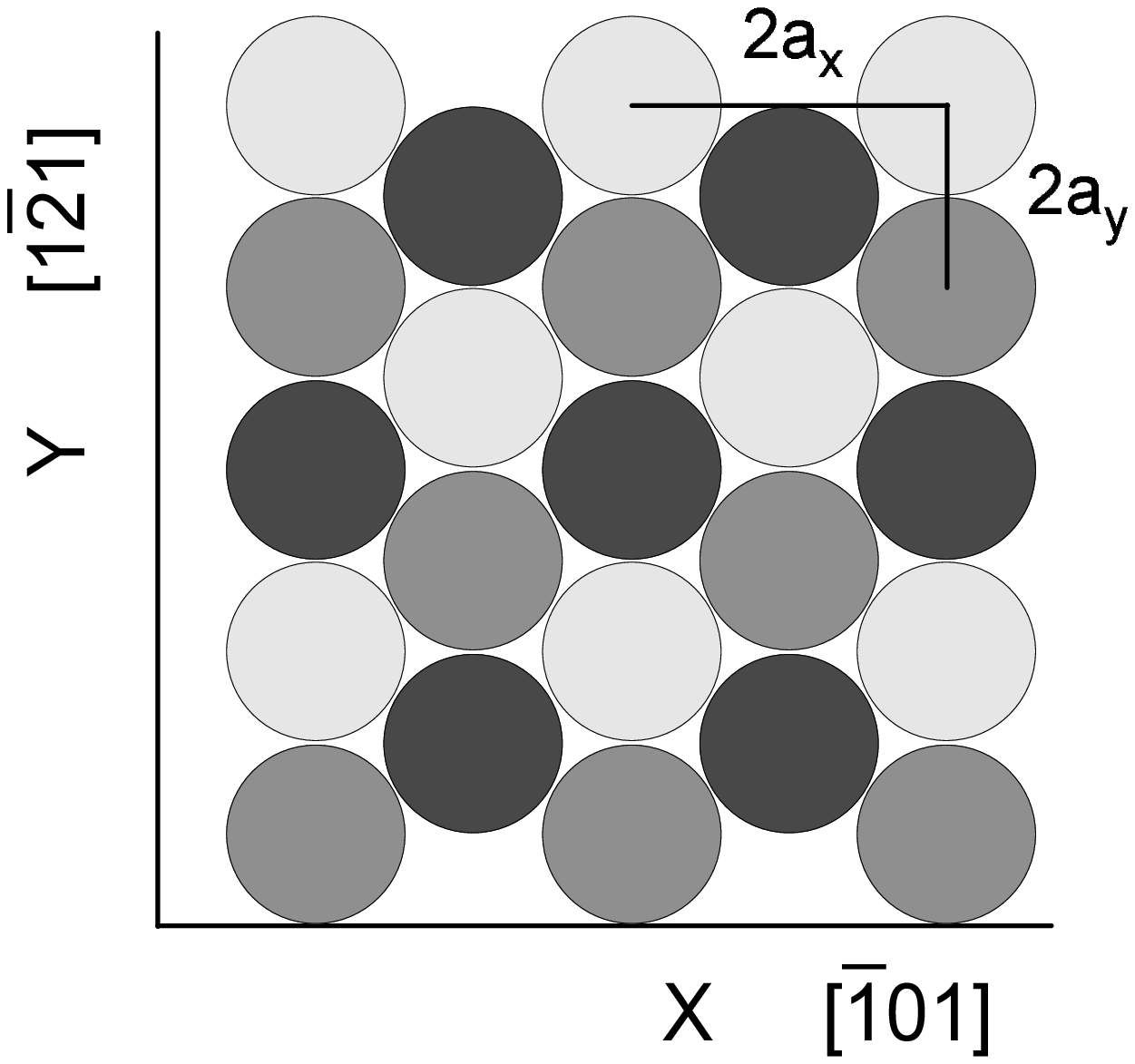}
\caption{ Schematic top view of the bcc(111) surface. Atoms from
three successive geometrical layers represent positions of columns
in the SOS model. The Z axis is normal to the (111) plane and
$a_x=a\sqrt{2}/2$, $a_y=a\sqrt{6}/6$.} \label{figure1}
\end{figure}
To study  an adsorbate-induced faceting on a spherical surface we
use a solid-on-solid model of Ref.~\cite{oleksy} that describes
faceting of bcc(111) surface. Let us recall the main assumptions.
Atoms along the closed packed direction [111] form columns which
are represented in the SOS model  by their positions on the (111)
plane, $i=(i_x,i_y)$,  and  heights, $h_i$. Column positions form
a triangular lattice (see Fig.\ \ref{figure1}) which is obtained
by projection of the bcc crystal lattice on the (111) plane. Each
column comprises  substrate atoms and one adsorbate atom at the
top. This implies a constant coverage of  1 physical monolayer.
There is also a restriction imposed on column heights: the nearest
neighbour column heights can  differ only by $\pm 1, \pm 2$. A
surface atom represented in the model by $(i_x,i_y,h_i)$ has a
position $(i_x a_x,i_y a_y,h_i a_z)$ in the bcc crystal, where
$a_z$ is the distance between neighbouring (111) layers
($a_z=a\sqrt{3}/6$) and {\em a} is the  lattice constant. The
model Hamiltonian represents surface formation energy which
depends on column heights $h_i$ in the following form
\begin{widetext}
\begin{eqnarray}\label{hsos}
H =&&
  \frac{1}{2}\sum\limits_i
    \left\{
      \sum\limits_{j_1}
         {\left[
           J_1 \delta
             \left(
                 \left| h_i  - h_{j_1 } \right| - 1
             \right)
             + K_1 \delta
             \left(
                 \left| h_i  - h_{j_1 }\right|-2\right)
             \right]
         }  \right.
     + \sum\limits_{j_2}
           {\left[
              2 J_2 \delta
                \left(
                  \left| h_i  - h_{j_2 }\right|
                \right) + \left( 2 J_2  + K_2\right)  \delta
                \left(
                  \left| h_i  - h_{j_2 } \right| - 3
                \right)
           \right] }\nonumber\\
   & &
     +
     \left.
       J_2 \sum\limits_{j_3}
         {\left[
               \delta
               \left( \left| h_i  - h_{j_3 } \right| - 2 \right)
               + \delta
               \left(
                  \left| h_i  - h_{j_3}\right| - 4
               \right)
         \right]}
     \right\}+ N J_0,
\end{eqnarray}
\end{widetext}

where the sums over $j_1$, $j_2$, and $j_3$ represent the sums
over first, second, and third neighbours of the column at site
$i$, respectively. Model parameters $J_0$, $J_1$,  $J_2$,  $K_1$,
and $K_2$ can be expressed by interaction energies between
substrate and adsorbate atoms (for details see \cite{oleksy}).

\subsection{Initial conditions for spherical surfaces }\label{s2i}

\begin{figure}
\centering
\includegraphics[width=7.5cm]{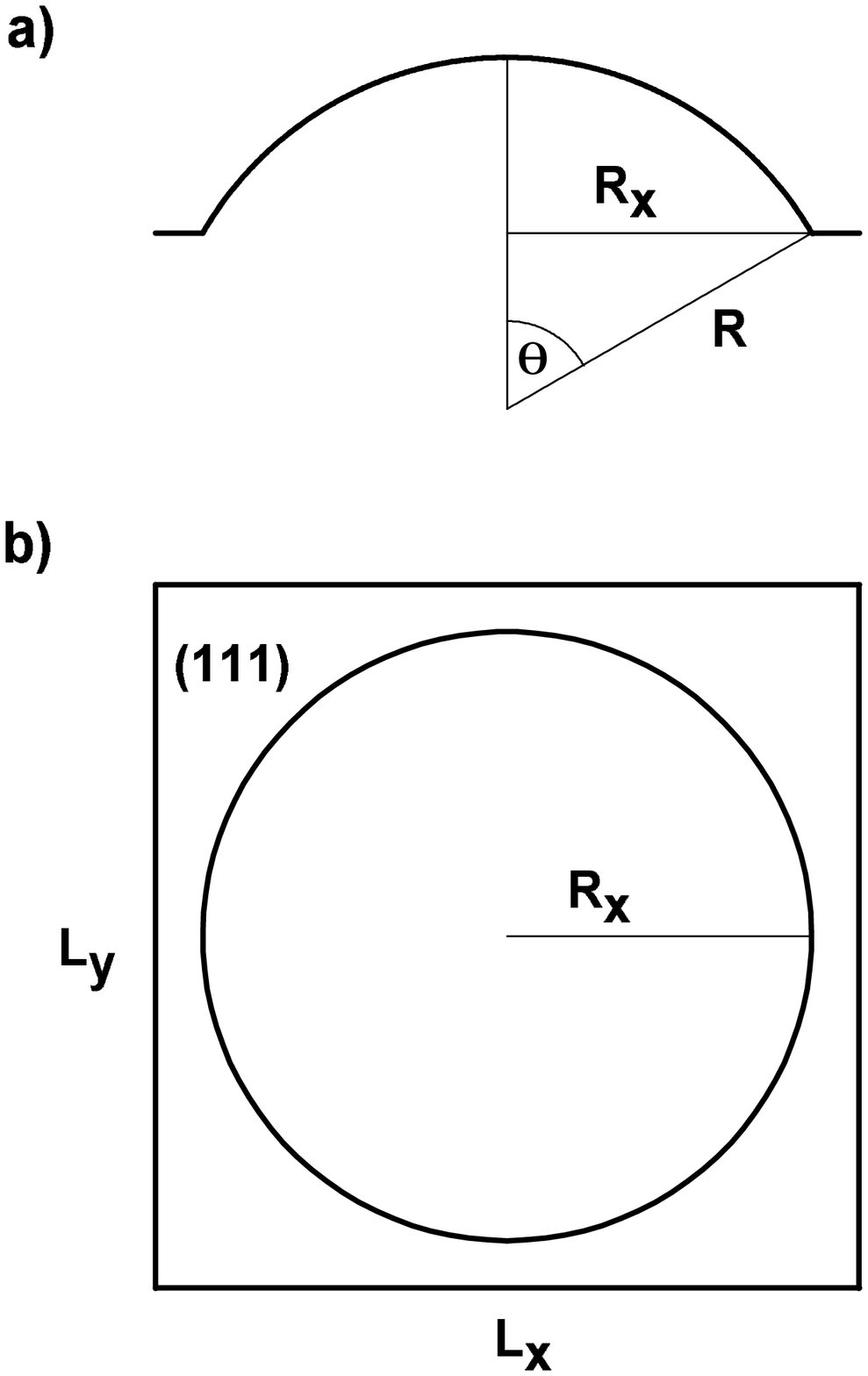}
\caption{  Schematic (a) side view (b) top view of a spherical
surface.} \label{figure2}
\end{figure}

\begin{figure}
\centering
\includegraphics[width=7.5cm]{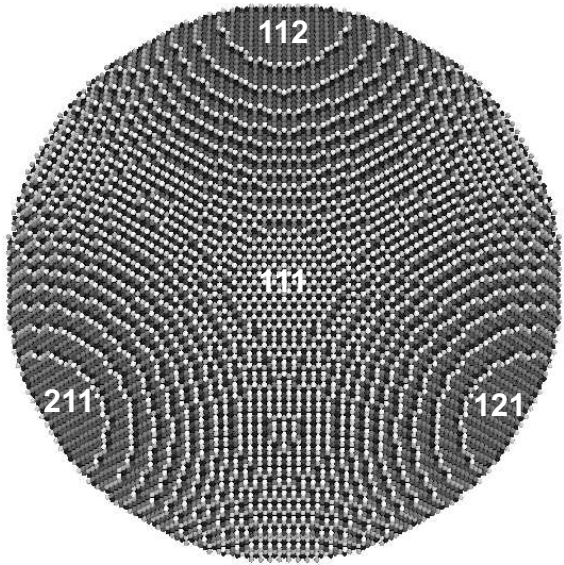}
\caption{Top view of a spherical  surface. The color of a surface
atom represents the number of its nearest neighbours (NN) with
smaller heights. The brightest color is  for 6 NN  and the darkest
one for 0 NN. } \label{sphere}
\end{figure}

We assume that the surface has a initially spherical shape
determined by the radius $R$ and the angular radius $\theta$ (see
Fig.\ \ref{figure2}). The integer values of column heights, $h_i$,
that minimize a deviation from the perfect  spherical shape, are
chosen. Positions of columns on the (111) plane form  a circle of
radius $R_x=R \sin\theta$. We choose fixed boundary conditions,
i.e., outside the circle  the columns form the (111) face and
their heights are frozen. Such boundary conditions make
calculations much easier because the number of interaction
constants in Hamiltonian (\ref{hsos}) can be reduced  to one, as
will be shown in what follows. It is worth noting that we can also
study a flat (111) surface by an appropriate choice of initial
column heights in the circle.  In this way we can compare
adsorbate-induced faceting of flat and curved surfaces
\cite{szczep05b} using the same conditions, i.e. the annealing
temperature, annealing Monte Carlo time, the number of surface
atoms, and the boundary conditions.
The restriction imposed on column heights of the nearest
neighbours  (see Sec.\ \ref{s2}) limits the angular radius to
$\theta\le 20^\circ$. An example of a spherical surface with
$\theta=20^\circ$  is shown in Fig.\ \ref{sphere}, where a (111)
facet and \{211\} facets occur together with terraces. There are
also smaller facets with higher Miller indices. It is worth noting
that in all surface images shown in this paper, the color of a
surface atom represents the number of its nearest neighbours with
smaller heights. Such representation allows an easy identification
of different facets, e.g.  \{110\} facets are represented by value
3 and row, trough of \{211\} facets by 5,1, respectively.
Investigation of faceting of curved tungsten surface induced by
palladium \cite{szczep02} has revealed that step-like microfacets
are formed in (111) crystal zones. Hence, the present model will
be used to study faceting around the single [111] pole on a
spherical surface.

\subsection{Energy of spherical surface }\label{s2e}

 In Ref.~\cite{oleksy} it has
been shown that energies per column of ideal faces (111), (211),
and (110) are the same for the interactions limited to the nearest
neighbours. It turns out that the same is true for the energy of a
spherical surface. We calculated  the number of bonds between
nearest neighbours: $b_1$ with $|\Delta h|=1$ and $b_2$ with
$|\Delta h|=2$ for hundreds of angles $\theta$ from the interval
$[0^\circ,20^\circ]$ and for constant number of columns $N$. For
most of the angles,  $b_1$ and $b_2$  are exactly the same as in
the case of the ideal (111) surface ($\theta=0^\circ$). For a few
angles the values of $b_1$ and $b_2$ are a little changed e.g.,
$|\Delta b_1|/b_1=3.8\times 10^{-4}$ for $R_x=93 a_x$ ($N=23497$)
and $|\Delta b_1|/b_1=5.6\times 10^{-5}$ for $R_x=370 a_x$
($N=327367$), and such angles will be omitted. On the other hand,
permitted changes of a configuration conserve $b_1$ and $b_2$.
Thus the energy of the nearest neighbours interactions is
conserved in this SOS model and it can be neglected (or treated as
the reference energy). For this reason we can limit the number of
the model parameters to two ($J_2$, $K_2$). Similarly as in
\cite{oleksy}, we can eliminate one more parameter by choosing
$J_2$ as the unit of energy. We will work with dimensionless
quantities : energy $\tilde H = H/J_2$, temperature $\tilde T =
k_BT/J_2$, and interaction constant $K = K_2/J_2$ (in what
follows, the tilde will be omitted). It has been shown that the
energy of the  (211) face  is minimal when $-2<K<0$, whereas the
(111) surface is stable for $K>0$. Therefore simulations of a flat
SOS model \cite{oleksy}  with negative K (e.g. $K=-1.25$) show
faceting of a bcc(111) surface. In the present model, energy of a
spherical surface $E_{sp}$ decreases with $\theta$ for negative
$K$ but is still greater than the energy of a (211) face, e.g. the
energy per column for $K=-1.25$ are: $E_{111} =7$,
$E_{sp}(\theta=20^\circ)=6.26$ and $E_{211} =5.75$. Thus we expect
that a spherical surface may undergo faceting after annealing at
an elevated temperature. In this paper we performed all
calculations for $K=-1.25$ for two reasons. First, the same value
was already used in faceting of a bcc(111) surface \cite{oleksy},
so one can compare results for two different models. Second, we
estimated the value of $K$ for the  Pd on W(111) system using
surface formation energies from Fig.~3 in Ref.~\cite{mad99b} and
the relations between $E_{hkl}$ and the model parameters
\cite{oleksy}. As the first-principle calculations \cite{mad99b}
were performed both with the local-density approximation (LDA) and
with the generalized gradient approximation (GGA), we obtained two
different values of $K$: $-1.06$ with LDA and $-1.25$ with GGA. We
finally chose the latter because the GGA is considered as  more
reliable than LDA \cite{mad99b}.

\section{Monte Carlo simulations}\label{s_MC}

 Properties of the spherical SOS model with constant
coverage are investigated by Monte Carlo (MC) simulation in
canonical ensemble using standard Metropolis algorithm
\cite{dplandau,binder88}. A new configuration  is generated  by
choosing two lattice sites \emph{i} and \emph{j} and changing
heights of columns at these sites: $(h_i, h_j)\rightarrow (h_i-3,
h_j+3)$. This is equivalent to  moving a substrate atom from a
site \emph{i} to a site \emph{j}. The new configuration is
accepted with probability $p=\min(1,\exp(-\delta E/kT)$, where
$\delta E$ is the energy change of the system generated by
transition to a new configuration. In the simulations we measure
the mean energy $\langle H\rangle$, the heat capacity
\begin{equation}
C_V=(\langle H^2\rangle -\langle H\rangle^2)/k_BT^2,
\end{equation}

and the mean-square  width of the surface,
\begin{equation}
{\delta h}^2 =
  \left<
    \frac{1}{Na^2}
      \sum\limits_{j}
         {
             \left( h_j  - \left<{h}\right>\right)^2
         }
    \right>.
\end{equation}
The latter quantity   depends also on the curvature radius,
therefore we introduce  the mean-square radial deviation
\begin{equation}
{\delta r}^2 =
  \left<
    \frac{1}{N a^2}
      \sum\limits_{j}
         {
             \left( r_j  - R\right)^2
         }
    \right>
\end{equation}
where  $r_j $ is distance of a  surface atom \emph{j} form the
center of the sphere.

\subsection{Faceting of curved surfaces}\label{s3}

In this section we investigate changes of a surface covered by a
physical monolayer of adsorbate  during the  warming up and
cooling down  processes. Simulations start at a temperature $T_0$,
with the spherical surface determined by the radius $R$ and the
angle $\theta$ (see section \ref{s2i}). The warming up process is
characterized by gradual elevation of temperature, $T_i=T_{i-1} +
\Delta T$. At each temperature the system is annealed for time
$\tau$, measured in Monte Carlo steps per site (MCS). After
reaching the maximal temperature, the system starts to cool down
by changing $\Delta T \rightarrow -\Delta T$. In order to measure
the temperature dependence of interesting quantities we choose
1000 configurations in the final stage at each temperature and we
repeat the  warming up -- cooling down cycle for 20 samples. In
what follows we will explain why averaging over samples is
important in this problem.
\begin{figure}
\centering
\includegraphics[width=7.5cm]{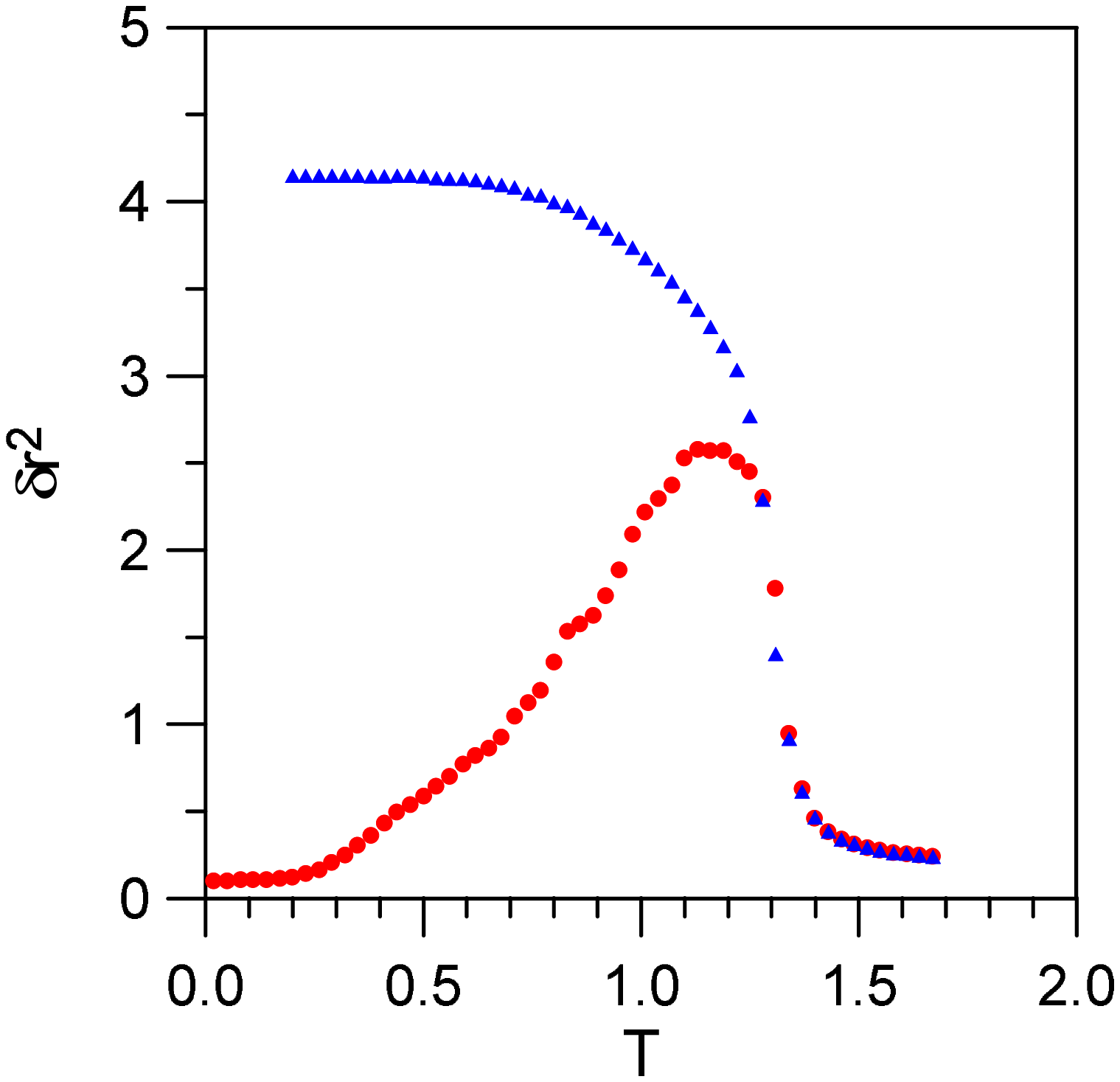}
\caption{ Temperature dependence of ${\delta r}^2$  during the
warming up  (red circles) and the cooling down (blue triangles) of
the system. } \label{figure3}
\end{figure}

\begin{figure*}
\centering
\includegraphics[width=14cm]{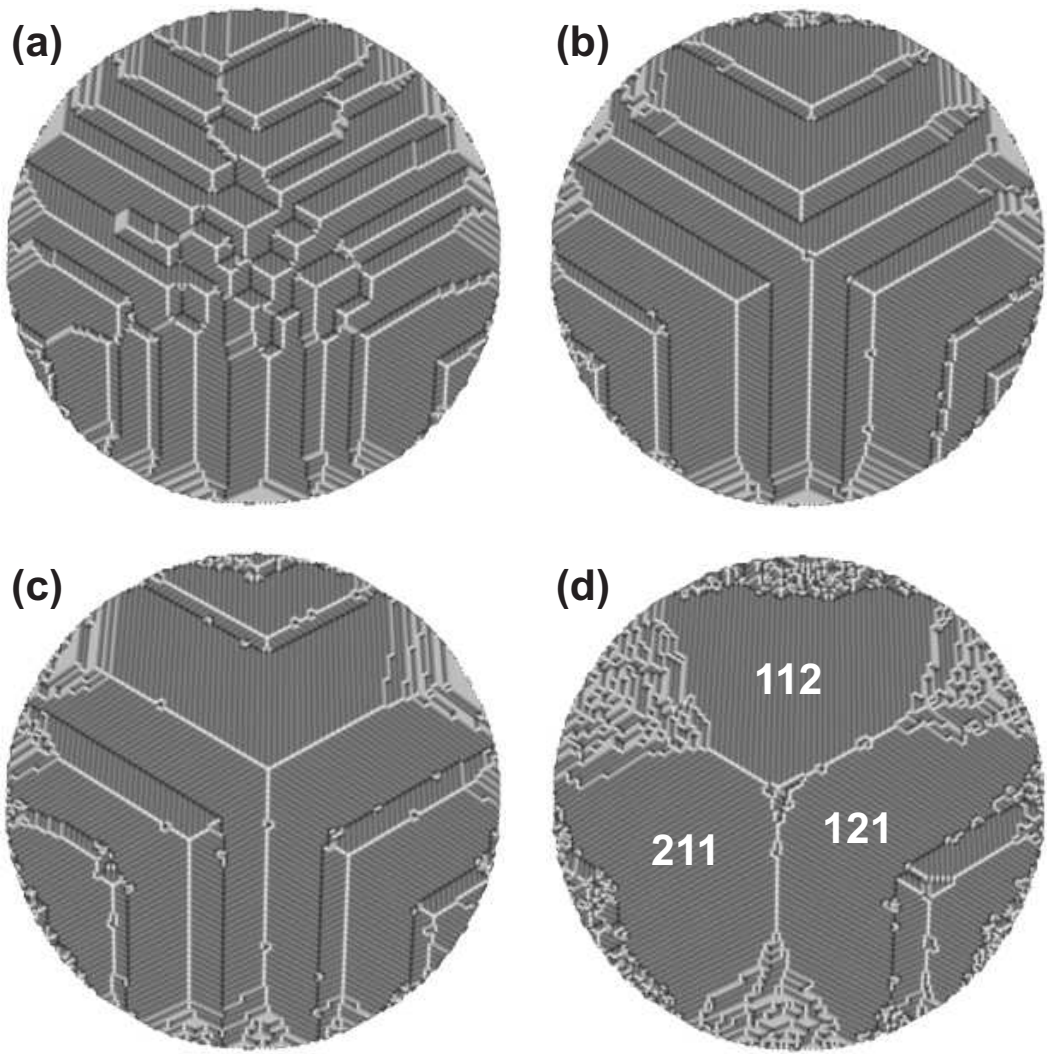}
\caption{Top view of the surface after annealing to: (a) $T=0.41$,
(b) $T=0.68$, (c) $T=0.86$ and (d) $T=1.10$. } \label{pic1}
\end{figure*}

\begin{figure}
\centering
\includegraphics[width=7.5 cm]{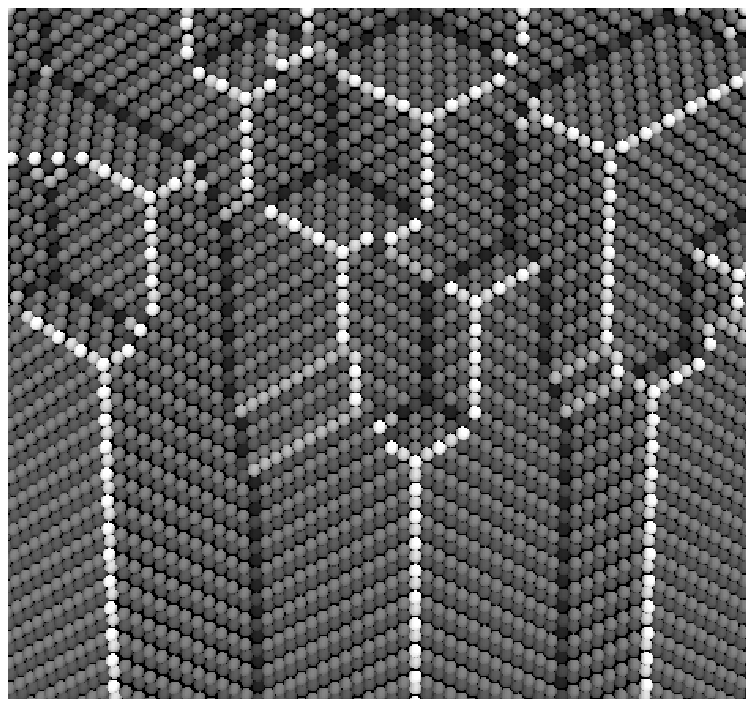}
\caption{Enlargement of the central  region from
Fig.~\ref{pic1}(a) where step-like facets and pyramids (pits)
coexist. } \label{pic_d}
\end{figure}

Results of simulations carried out for $K = -1.25$, $R_x=93 a_x$,
$\theta=20^\circ$, $\Delta T=0.03$ and $\tau= 2\times 10^5$ are
presented in Figs.~\ref{figure3}~--~\ref{figure5}. The temperature
dependence of ${\delta r}^2$ in the warming up -- cooling down
cycle clearly shows (see Fig.\ \ref{figure3}) that this process is
irreversible. For $T>0.3$ the mean-square radial deviation
increases as the temperature is elevated, which indicates that
\{211\} facets are formed on initially spherical surface. First we
observe two types of facets on the surfaces (see
Fig.~\ref{pic1}(a) and Fig.~\ref{pic_d}): 3-sided pyramids and
pits located near the pole, and elongated facets forming  steps
between the \{211\} faces.
The steps  are built of two alternating facets, e.g. (211) and
(121). They have characteristic parallel long edges, oriented
along $\langle311\rangle$ directions, that can be seen in FIM
experiments \cite{szczep02,szczep04} as bright lines. Note that
there are also concave edges (dark lines on Fig.~\ref{pic_d})
which are not present in FIM images. Further increase of the
temperature makes the   pyramids to disappear and reduces  the
number of steps (see for example Fig.~\ref{pic1}(b)). In other
words, a single pyramid with multiple edges is observed. As
temperature is increased, even more,  the number of these edges is
reduced (see Fig.~\ref{pic1}(c)). At high enough temperature, it
is possible to observe a single 3-sided pyramid on the surface
(see Fig.\ \ref{pic1}(d)); sometimes this pyramid has a double
edge. Above $T_d \approx 1.31$, a rapid decline of  ${\delta r}^2$
is observed. This  indicates a defaceting transition at which the
pyramid disappears (see e.g. Fig.~\ref{pic2}(a)). Although
${\delta r}^2$ is very small above $T_d$,  the surface is not
spherical. We observe also that the surface energy rapidly
increases at $T_d$ and the heat capacity attains its maximum. A
similar defaceting transition was obtained in simulations of
faceting of a bcc(111) surface \cite{oleksy} and observed in LEED
experiments \cite{song95} for Pd on Mo(111). The reverse process
of slow cooling down the system from the highest temperature,
leads again to faceting at the temperature $T_d$. Hence this is a
reversible phase transition.

\subsection{Measurements after fast cooling}\label{s3fc}

In FIM experiment \cite{szczep04}  the number of parallel edges on
a faceted curved surface was measured as a function of the
annealed temperature. Because  FIM images are taken after a rapid
cooling to a low temperature, we adopted a similar procedure to
measure the average number of parallel edges, $n_e$.
Configurations chosen at a temperature $T$ were quickly cooled
down (the system was held for 20 MCS at each $T_i$) to the
temperature $T_m= 0.2$ at which calculations where performed. This
procedure reduces the thermal disorder and makes it possible to
use a simple  algorithm for  edges searching.
\begin{figure}
\centering
\includegraphics[width=7.5cm]{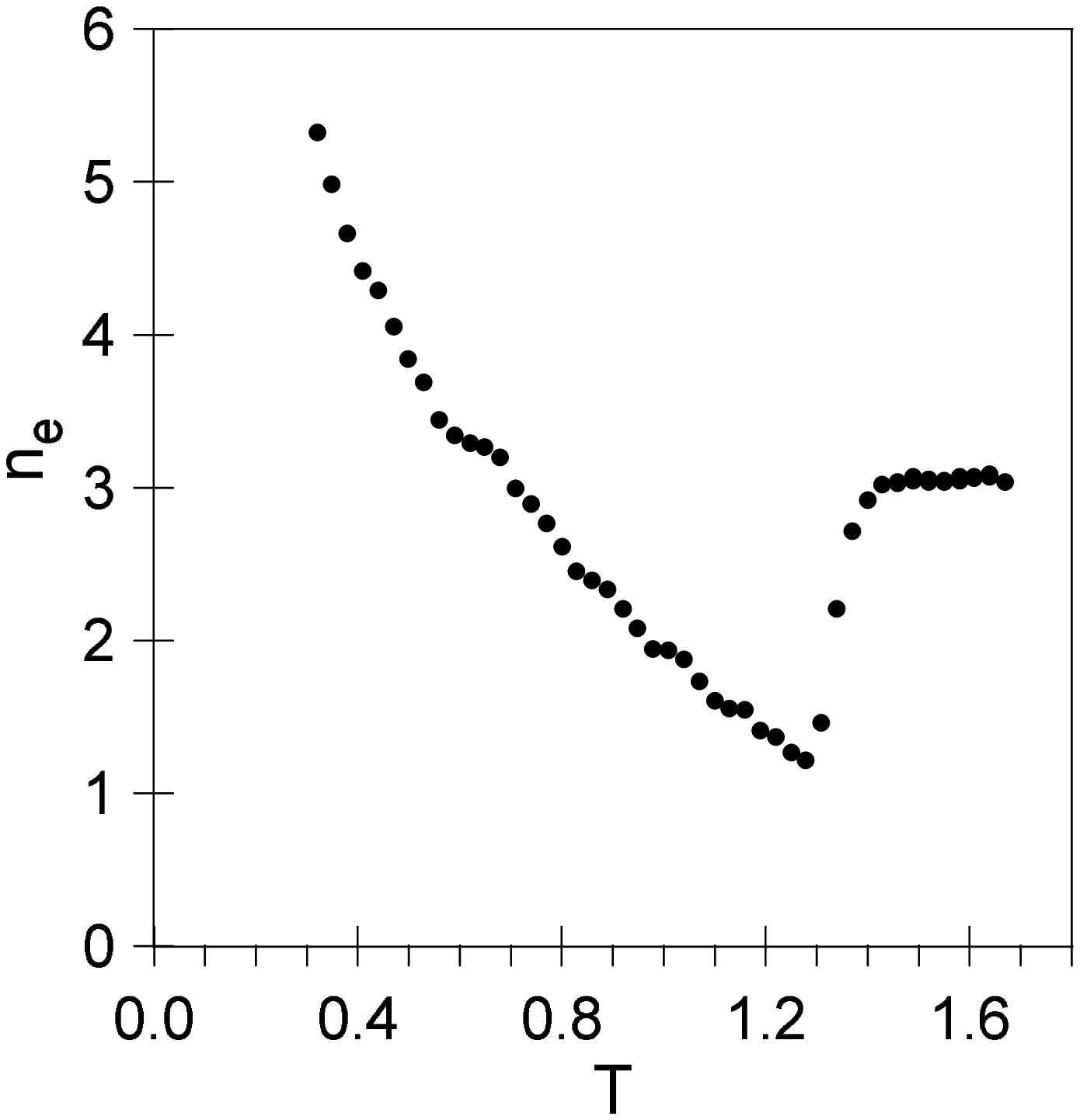}
\caption{The average number of parallel edges versus temperature
calculated  after fast cooling to $T= 0.2$.} \label{figure4}
\end{figure}
It is interesting to note that the number of parallel  edges
decreases with the temperature  down to the  defaceting
temperature and then suddenly increases (see Fig.~\ref{figure4}).
Similar behavior was observed in the FIM experiment
\cite{szczep04}. The reason why fast cooling can reproduce a
pyramid with multiple edges lays in morphology of disordered
surface. When the temperature is elevated over $T_d$, the pyramid
is initially damaged near the apex and along the edges. In these
regions the surface is not spherical but has a disordered
hill-and-valley structure, i.e. there are many  small \{211\}
facets which form irregular, short \{211\} steps. As the
temperature is elevated, the disordered regions broaden (see
Fig.~\ref{pic2}(a)).
\begin{figure} \centering
\includegraphics[width=7.5cm]{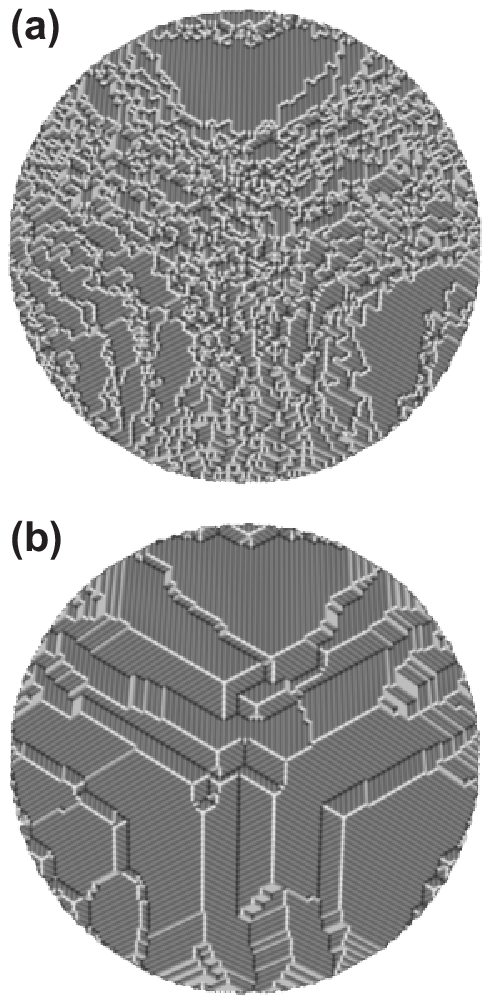}
\caption{Snapshot of the surface (a) at $T=1.4$, (b) after fast
cooling from $T=1.4$  to $T=0.2$. } \label{pic2}
\end{figure}
On the other hand, fast cooling causes the system to return to a
pyramid with multiple edges  (see Fig.~\ref{pic2}(b)) and
multiplicity depends on the width of the  disordered region and
the cooling rate.  This fact can be used to estimate $T_d$ from
experimental results, provided that the  temperature obtained is
below the desorption temperature. We have also checked how this
effect depends on the cooling rate. If the cooling is 10 times
faster, then this effect does not occur, i.e. the disordered
surfaces remains frozen. In contrast to this, cooling of the
disordered phase will result in   a big 3-sided pyramid on the
surface for  cooling rates over hundred times slower.
\begin{figure}
\centering
\includegraphics[width=7.5cm]{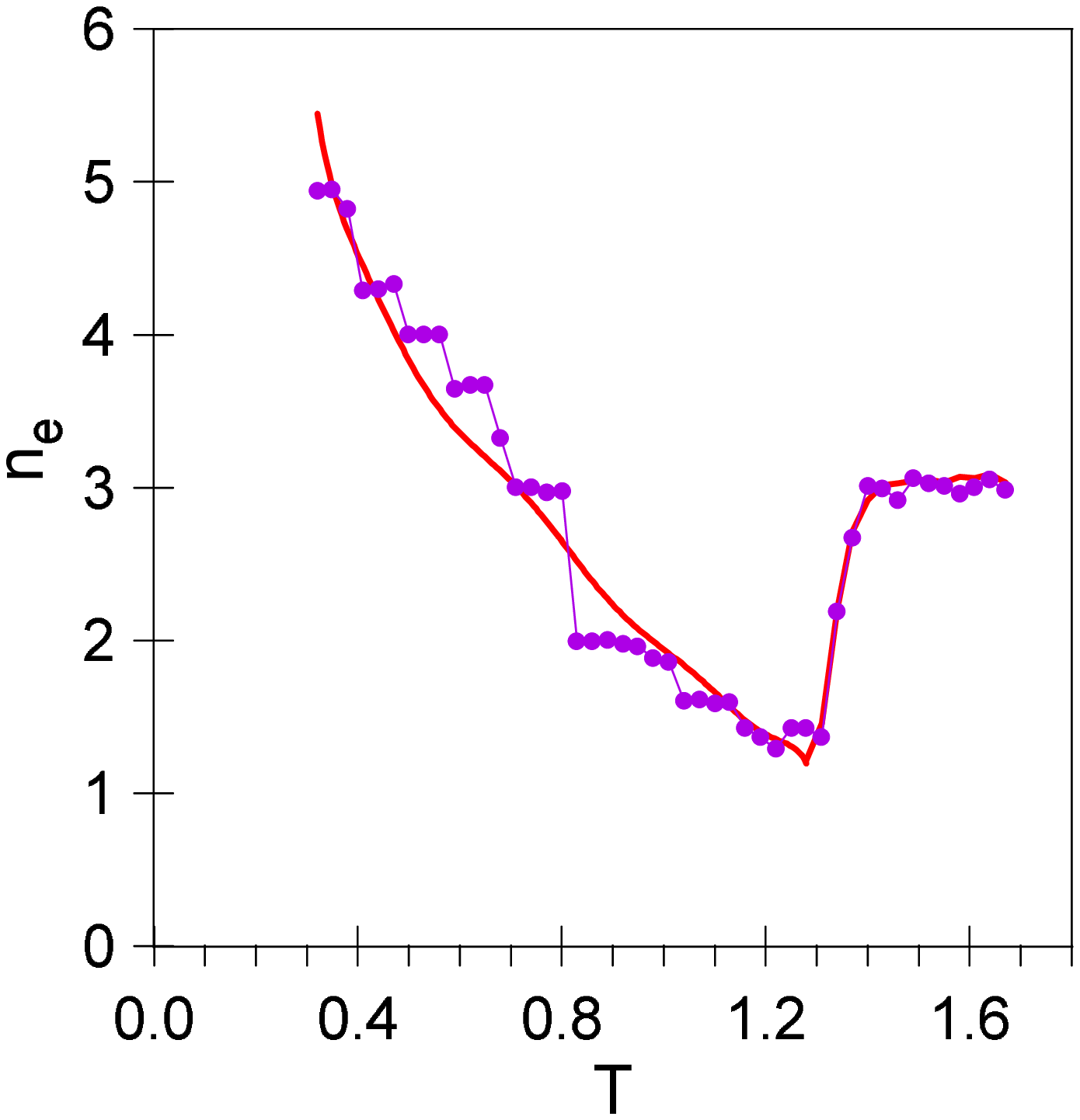}
\caption{ The number of parallel edges  for a single sample
(circles) and averaged over 20 samples (the thick line represents
data depicted in Fig.\ref{figure4}). } \label{figure5}
\end{figure}

The quantities like $n_e$, ${\delta r}^2$, ${\delta h}^2$ have
different temperature behavior for a  single sample and when they
are averaged over many samples. The single sample temperature
dependence of these quantities have step-like character (e.g. see
Fig.~\ref{figure5}). We think that this effect is connected with
large sizes of step-like  \{211\} facets on a curved surface. A
massive transport of atoms is needed to reduce one convex edge by
merging two neighbouring \{211\} steps. Of course, there should be
an energy barrier for such process and its height must depend on
the size of \{211\} facets. Therefore the width of the plateau
occurring in $n_e(T)$ in Fig.~\ref{figure5} increases with the
annealing temperature. On the other hand, $n_e(T)$ averaged over
many sample becomes a smooth function of temperature because the
growth of facets  proceeds in individual way for each sample due
to the fact that the distances between edges are not constant at a
given T (see Fig.~\ref{pic1}), but they have visible dispersion.

\subsection{Defaceting of curved surfaces}\label{s3d}

\begin{figure}
\centering
\includegraphics[width=7.5cm]{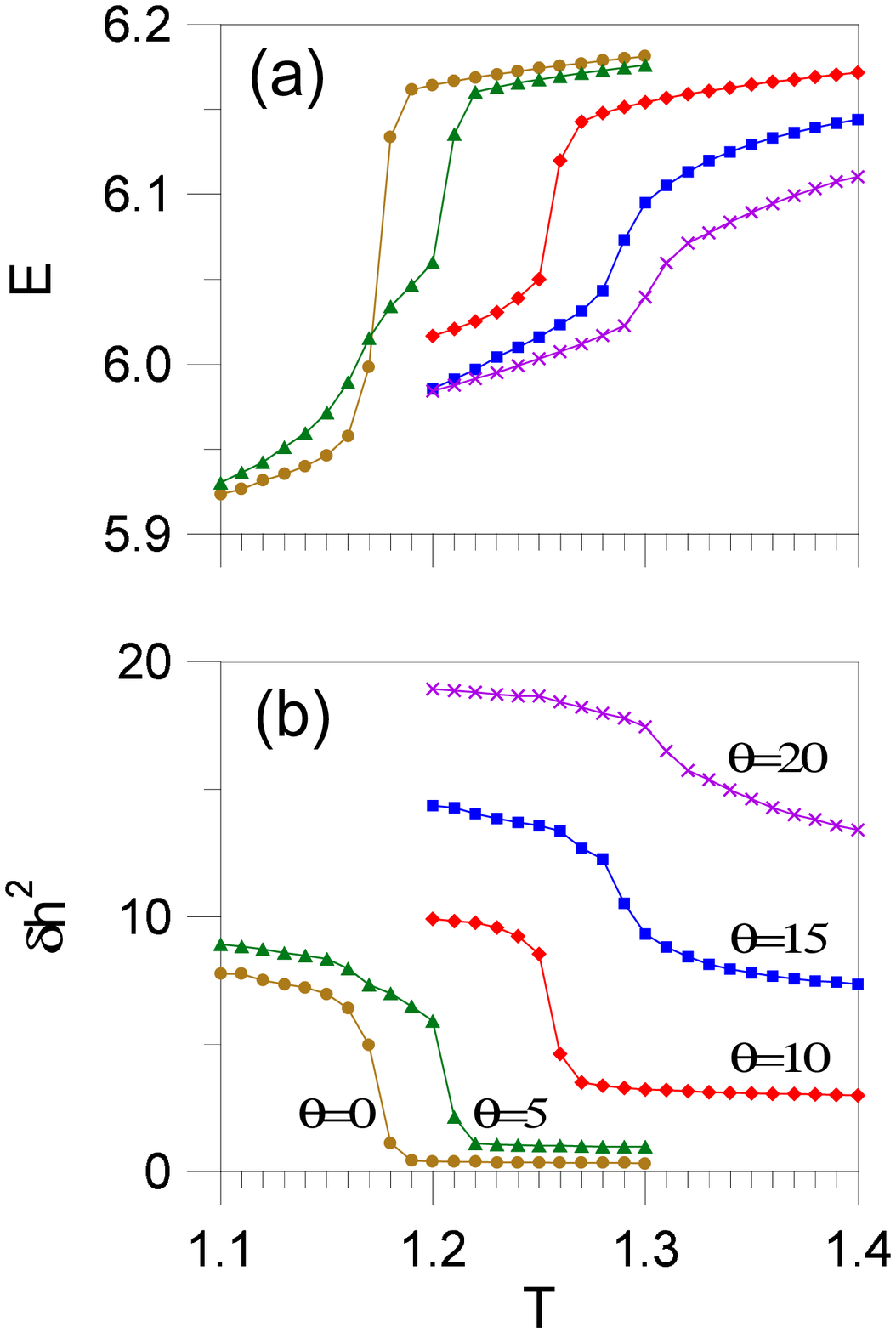}
\caption{Dependence of (a) surface energy, (b) square mean width
on temperature near faceting - defaceting transition for several
angles $\theta$. } \label{figure6}
\end{figure}

It has been mentioned in Sec \ref{s3} that defaceting temperature,
$T_d$, of spherical surface with ($\theta=20^\circ$) is greater
than in the case of a flat surface. To investigate the dependence
of $T_d$ on the surface curvature we performed a series of
simulations for several angles $\theta$, keeping the number of
columns $N$ (or $R_x$) fixed. This means that the radius $R=R_x/
\sin\theta$ of a spherical surface was varied. Simulation started
from a high temperature and as the system was  cooled down with
$\Delta T=0.01$, we measured the surface energy, the heat
capacity, the mean-square width and the mean-square radial
deviation near the reversible phase transition.  The system was
held at each temperature for $10^6$ MCS. We see in
Fig.\ref{figure6}(a) that the  surface energy of a flat surface
rapidly decreases at the defaceting temperature $T_d$. This is the
first-order phase transition and the energy difference is equal to
the latent heat. Behaviors of $E$ and ${\delta h}^2$ indicate that
this phase transition differs from the roughening phase transition
\cite{tosat96,salanon88,nijs89,jaszczak97} which belongs to the
universality class of Kosterlitz-Thouless transition.

It is clearly shown in Fig.\ref{figure6} that the temperature of
defaceting transition increases as the  curvature of the surface
increases. For a surface with $\theta=20^\circ$ the change of
$T_d$ is about $10\%$ with respect to the flat case. We also see
that the change of the surface energy at the phase transition
decreases with $\theta$. The same behavior is observed in the
dependence of ${\delta h}^2$ on $T$ and $\theta$ (see
Fig.\ref{figure6}(b)).\par We think that the increase of $T_d$
with $\theta$ is related to the entropy reduction  in the
disordered phase. The surface free energy (SFE), $F=E-T S$, of the
faceted phase $F_f$ at the transition temperature $T_d$ is equal
to SFE of the disordered phase $F_d$. Therefore, the temperature
of phase transition can be obtained from $T_d(\theta)=\Delta
E(\theta,T_d)/ \Delta S(\theta,T_d)$, where $ \Delta X=X_d - X_f$
for $X=E,S$. As is seen in Fig.~\ref{figure6}(a) the change of
internal energy $ \Delta E$ decreases with $\theta$ at $T_d$. Thus
the change of entropy $ \Delta S$ should also be a decreasing
function of $\theta$. The decrease of entropy in disordered phase
with $\theta$ can be justified  by the fact that one can observe
an increase of \{211\} areas and a contraction of disordered
regions on the surface at $T_d$ as curvature (or $\theta$)
increases.

\section{Discussion}

It is shown that a simple solid-on-solid model  can be applied to
study adsorbate-induced faceting of the $[111]$ pole region of a
spherical surface. Just as in the case of faceting of a bcc(111)
surface, it is possible to specify interaction constants that
describes a situation when an initially prepared spherical surface
covered by a physical monolayer of adsorbate has a surface
formation energy greater than \{211\} surfaces. Thus, the faceting
of a curved surface is thermodynamically favorable. Results of
Monte Carlo simulations show that the morphology of a faceted
surface depends on the annealing temperature. One can distinguish
three stages: (1) Formation of 3-sided pyramids around the [111]
pole and step-like \{211\} facets between the faces of  \{211\}
orientations. The latter form long edges observed in FIM
experiments \cite{szczep02,szczep04}. (2) Disappearing of pyramids
around the [111] pole and domination of step-like facets whose
number decreases with the temperature. (3) Formation of a single
large pyramid on the surface. Note that the last two cases were
observed in  FIM experiments. The first case, a mixture of two
kinds of facets, has not been experimentally detected yet. Using
our model, we can investigate in detail the structure of a faceted
spherical surface. In particular, it is possible to examine
properties difficult for experimental studies: concave edges
between elongated facets, transition from step-like facets to
pyramids and pits, defects, etc.

It is shown that there exists a temperature of faceting --
defaceting transition $T_d$ above which disordered regions appear
along pyramid edges. Although mean-square radial  deviation
${\delta r}^2$ rapidly decreases above $T_d$, the surface is not
spherical. Antczak et al \cite{antczak01} observed in a FEM
experiment of Pt on W tip emitter that faceting diminishes at
$T>1600$K and the hemispherical form of the emitter is recovered.
As platinum is desorbed from tungsten emitter at temperatures
above 1900K \cite{antczak01}, this observation can indicate that
defaceting occurs in Pt on W system. It would be interesting to
verify experimentally this suggestion.  We found that at
temperatures just above $T_d$, disordered regions are not
"spherical" but they have disordered hill-and valley structure
with many small \{211\} facets which form irregular, short \{211\}
steps. This fact seems to be important for measurements of high--
temperature properties performed after fast cooling down to a low
temperature, a situation typical e.g. in FIM experiments. It is
shown that fast cooling may transform the disordered surface into
a multi--step surface (see Fig.\ref{pic2}(b)). This is manifested
in a temperature behavior of the number of parallel edges $n_e$,
which exhibits an unexpected increase  above the phase transition
temperature $T_d$ (see Fig.\ref{figure4}). Similar behavior of
$n_e$ has been found  for oxygen adsorbed on curved W
\cite{szczep04}. We believe that this fact can be used to
determine the temperature of defaceting, provided that $T_d$ is
below the desorption temperature (desorption is not included in
the model). Formation of facets after a rapid cooling of
disordered surface can be explained by a fast massive atomic
rearrangement. Such effect has been observed in a Pd/Mo(111)
system \cite{song95} in which 3-sided pyramidal facets occur on
the bcc(111) surface after annealing to higher temperatures. Above
$T_d\approx 870$K the pyramids disappear and surface becomes flat.
Attempts to freeze a high -- temperature flat phase ended up with
a faceted surface even for 60 K/s of cooling rate.

Another interesting property of faceting found in the present  MC
simulation is the dependence of $T_d$ on the surface curvature.
The temperature of transition $T_d$ on a flat surface is smaller
than $T_d$ on a curved surface. In FEM experiment
\cite{antczak01}, faceting of a Mo emitter covered by Pd has been
observed  in temperature range $750 - 1040$K, so at temperatures
above $T_d\approx 870K$ for the Pd/Mo(111) system \cite{song95}.
However, this can not be regarded as the evidence for higher
defaceting temperature in a curved system because FEM observations
were made at room temperature and the cooling effect could affect
these FEM results.

\subsection*{Acknowledgements}
We would like to thank  dr Andrzej Szczepkowicz for  discussions
and help in preparing images.

\end{document}